\documentclass[12pt]{article}
\usepackage{times}
\usepackage{amsfonts}
\begin{document}
\baselineskip = 0.60 true cm \thispagestyle{empty} \vspace{1.8cm}

{\bf Phase operators, phase states and vector phase states for $SU_3$ and $SU_{2,1}$}

\vspace{1.5cm}

{\bf M. Daoud$^{1,2,3,4}$ and M. R. Kibler$^{1,2,3}$}

\vspace{0.8cm}

\noindent$^1$ Universit\'e de Lyon, 69361 Lyon, France \\ 
$^2$ Universit\'e Claude Bernard Lyon 1, 69622 Villeurbanne, France \\ 
$^3$ CNRS/IN2P3, Institut de Physique Nucl\'eaire, 69622 Villeurbanne, France \\ 
$^4$ D\'epartement de Physique, Facult\'e des Sciences, Agadir, Morocco

\vspace{0.5cm} 
\noindent E-mail : m\_daoud@hotmail.com and m.kibler@ipnl.in2p3.fr

\vspace{1.5cm} 

{\bf Abstract} 

\baselineskip=18pt
\medskip

This paper focuses on phase operators, phase states and vector phase states for the $sl_3$ Lie algebra. We introduce 
a one-parameter generalized oscillator algebra ${\cal A}_{\kappa}(2)$ which provides a unified scheme for dealing with 
$su_3$ (for $\kappa < 0$), $su_{2,1}$ (for $\kappa > 0$) and $h_4 \otimes h_4$ (for $\kappa = 0$) symmetries. Finite- 
and infinite-dimensional representations of ${\cal A}_{\kappa}(2)$ are constructed for $\kappa < 0$ and $\kappa \geq 0$, 
respectively. Phase operators associated with ${\cal A}_{\kappa}(2)$ are defined and temporally stable phase states 
(as well as vector phase states) are constructed as eigenstates of these operators. Finally, we discuss a relation 
between quantized phase states and a quadratic discrete Fourier transform and show how to use these states for 
constructing mutually unbiased bases.

\newpage

\section{INTRODUCTION AND MOTIVATIONS}

It is well known that defining an Hermitian (or unitary, by means of an exponentiation trick) phase operator 
for the Fock space of the isotropic harmonic oscillator and, more generally, for an infinite-dimensional 
Hilbert space is not an easy problem.$^{\cite{SusskindGlogower}}$ Pegg and Barnett$^{\cite{pegg-barnett}}$ solved 
it by replacing the oscillator algebra by a truncated oscillator algebra and thus were able to give a description 
of the phase properties of quantum states for the single modes of the electromagnetic field. In this spirit, 
Vourdas$^{\cite{Vourdas1990}}$ introduced phase operators and phase states (i.e., eigenvectors of a phase operator) 
for $su_2$ and $su_{1,1}$; for the $su_{1,1}$ Lie algebra, he noticed that the infinite-dimensional character of 
the representation space prevents to define a unitary phase operator. Phase operators and phase states for other 
symmetries were also studied. In particular, Klimov {\em et al.}$^{\cite{klimov1}}$ obtained phase states for some 
specific representations of $su_{3}$. 

Recently, a generalized oscillator algebra ${\cal A}_{\kappa}$, depending on a real parameter $\kappa$, was introduced 
to cover the cases of Lie algebras $su_2$ (for $\kappa < 0$) and $su_{1,1}$ (for $\kappa > 0$) as well as Weyl-Heisenberg 
algebra $h_4$ (for $\kappa = 0$).$^{\cite{AKW,daoud-kibler}}$ Temporally stable phase states were defined 
as eigenstates of phase operators for finite-dimensional ($\kappa < 0$) and infinite-dimensional 
representations ($\kappa \geq 0$) of the ${\cal A}_{\kappa}$ algebra.$^{\cite{daoud-kibler}}$ In the finite-dimensional case, corresponding 
either to $\kappa < 0$ or to $\kappa \geq 0$ with truncation, temporally stable phase states proved to be useful for 
deriving mutually unbiased bases.$^{\cite{AKW,daoud-kibler}}$ Such bases play an important role in quantum information 
and quantum cryptography. 

In this paper, we introduce an algebra, noted ${\cal A}_{\kappa}(2)$, which generalizes the ${\cal A}_{\kappa}$ algebra. For 
$\kappa < 0$, this new algebra is similar to that considered in the seminal work of Palev$^{\cite{palev1, palev2}}$ in the 
context of $A_n$-statistics. The ${\cal A}_{\kappa}(2)$ algebra allows to give an unified treatment of algebras 
$su_3$ (for $\kappa < 0$), $su_{2,1}$ (for $\kappa > 0$) and $h_4 \otimes h_4$ (for $\kappa = 0$). When we started 
this work, our aim was to study in an unified way: (i) phase operators for $su_3$, $su_{2,1}$ and $h_4 \otimes h_4$ and (ii) 
the corresponding phase states. We discovered, for $\kappa < 0$, that phase states can be defined only for partitions of the 
relevant Hilbert spaces and that a global definition of {\em phase states} requires the introduction of {\em vector phase states}, 
a concept that is closely related to that of vector coherent states. The notion of vector coherent states was strongly 
investigated by Hecht$^{\cite{Hecht}}$ and Zhang {\em et al.}$^{\cite{Gilmore}}$ at the end of the nineties. This notion was 
subsequently developed in Refs.~\cite{twareque2, twareque1, gazeau, twareque3} with applications to quantum dynamical systems 
presenting degeneracies. In particular, the authors of Ref.~\cite{twareque1} defined a vectorial generalization of the Gazeau-Klauder 
coherent states$^{\cite{gazeau}}$ leading to vector coherent states. Recently, this notion of vector coherent states was 
extensively investigated (see the works in Refs.~\cite{twareque2,twareque3}). 

This paper is organized as follows. The ${\cal A}_{\kappa}(2)$ generalized algebra is introduced in Section 2. We then define a 
quantum system associated with this algebra and generalizing the two-dimensional harmonic oscillator. In Section 3, phase
operators and temporally stable vector phase states for the ${\cal A}_{\kappa}(2)$ algebra with $\kappa < 0$ are constructed. The phase operators 
and the corresponding temporally stable phase states for ${\cal A}_{\kappa}(2)$ with $\kappa \geq 0$ are presented in Section 4. Section 5 
deals with a truncation of the ${\cal A}_{\kappa}(2)$ algebra, with $\kappa \geq 0$, necessary in order to get unitary phase operators. 
In Section 6, we show how a quantization of the temporality parameter occurring in the phase states for ${\cal A}_{\kappa}(2)$ with $\kappa < 0$ 
can lead to mutually unbiased bases. 

\section{GENERALIZED OSCILLATOR ALGEBRA ${\cal A}_{\kappa}(2)$}

\subsection{The algebra}

We first define the ${\cal A}_{\kappa}(2)$ algebra. This algebra is generated by six linear 
operators $a_i^-$, $a_i^+$ and $N_i$ with $i=1,2$ satisfying the commutation relations
			\begin{eqnarray}
[a_i^- , a^+_i] = I + \kappa (N_1 + N_2 + N_i), \quad 
[N_i , a_j^{\pm}] = {\pm} \delta_{i,j} a_i^{\pm}, \quad i,j = 1,2
			\label{commutation1}
			\end{eqnarray}
and
			\begin{eqnarray}
[a_i^{\pm} , a_j^{\pm}] = 0, \quad i \neq j,  
 			\label{commutation2}
			\end{eqnarray}
complemented by the triple relations
			\begin{eqnarray}
[a_i^{\pm} , [a_i^{\pm} , a_j^{\mp}]] = 0, \quad i \neq j. 
			\label{commutation3}
			\end{eqnarray}
In Eq.~(\ref{commutation1}), $I$ denotes the identity operator 
and $\kappa$ is a deformation parameter assumed to be real. 

Note that the ${\cal A}_{\kappa}$ algebra introduced in Ref.~\cite{daoud-kibler} 
formally follows from ${\cal A}_{\kappa}(2)$ by omitting the relation 
$[a_2^- , a^+_2] = I + \kappa (N_1 + 2 N_2)$ and by taking 
			\begin{eqnarray}
a_2^- = a_2^+ = N_2 = 0, \quad 
a_1^- = a^-, \quad
a_1^+ = a^+, \quad
N_1   = N
 			\nonumber
			\end{eqnarray}
in the remaining definitions of ${\cal A}_{\kappa}(2)$. Therefore, 
generalized oscillator algebra ${\cal A}_{\kappa}$ in Ref.~\cite{daoud-kibler} 
should logically be noted ${\cal A}_{\kappa}(1)$.

For $\kappa = 0$, the ${\cal A}_{0}(2)$ algebra is nothing but the 
algebra for a two-dimensional isotropic harmonic oscillator and thus 
corresponds to two commuting copies of the Weyl-Heisenberg algebra $h_4$. 

For $\kappa \not= 0$, the ${\cal A}_{\kappa}(2)$ algebra  
resembles the algebra associated with the so-called $A_n$--statistics (for $n=2$) which was introduced by 
Palev$^{\cite{palev1}}$ and further studied from the microscopic point of view by Palev and Van der 
Jeugt.$^{\cite{palev2}}$ In this respect, let us recall that $A_n$--statistics is described by the $sl_{n+1}$ Lie algebra 
generated by $n$ pairs of creation and annihilation operators (of the type of the $a_i^+$ and $a_i^-$ operators 
above) satisfying usual commutation relations and triple commutation relations. Such a presentation of $sl_{n+1}$ 
is along the lines of the Jacobson approach according to which the $A_n$ Lie algebra can be defined by means 
of $2n$, rather than $n(n+2)$, generators satisfying commutation relations and triple commutation 
relations.$^{\cite{jacobson}}$ These $2n$ Jacobson generators correspond to $n$ pairs of creation and annihilation 
operators. In our case, the ${\cal A}_{\kappa}(2)$ algebra for $\kappa \not= 0$, with two pairs of Jacobson 
generators ($(a_i^+, a_i^-)$ for $i = 1,2$), can be identified to the Lie algebras $su_3$ for 
$\kappa < 0$ and $su_{2,1}$ for $\kappa > 0$. This can be seen as follows. 

Let us define a new pair $(a_3^+ , a_3^-)$ of operators in terms of the two pairs $(a_1^+ , a_1^-)$ and 
$(a_2^+ , a_2^-)$ of creation and annihilation operators through
 	\begin{eqnarray}
a_3^+ = [a_2^+ , a_1^-], \quad a_3^- = [a_1^+ , a_2^-].
 	\label{lesdeuxa3}
 	\end{eqnarray}
Following the trick used in Ref.~\cite{AKW} for the ${\cal A}_{\kappa}(1)$ algebra, let us introduce 
the operators 
 	\begin{eqnarray}
& & E_{+\alpha} = \frac{1}{\sqrt{|\kappa|}} a_{\alpha}^+, \quad 
		E_{-\alpha} = \frac{1}{\sqrt{|\kappa|}} a_{\alpha}^-, \quad \alpha = 1,2,3 \nonumber \\
& & H_1 = \frac{1}{2 \kappa} [I + \kappa (2 N_1+ N_2)], \quad 
    H_2 = \frac{1}{2 \kappa} [I + \kappa (2 N_2+ N_1)] \nonumber
 	\end{eqnarray}
with $\kappa \not= 0$. It can be shown that the set $\{ E_{{\pm}\alpha} ; H_i : \alpha = 1,2,3 ; i = 1,2 \}$ 
spans $su_3$ for $\kappa < 0$ and $su_{2,1}$ for $\kappa > 0$.

\subsection{Representation of ${\cal A}_{\kappa}(2)$}

We now look for a Hilbertian representation of the ${\cal A}_{\kappa}(2)$ algebra  
on a Hilbert-Fock space ${\cal F}_{\kappa}$ of dimension $d$ with $d$ finite or infinite. Let 
 	\begin{eqnarray}
\{ \vert n_1 , n_2 \rangle : n_1, n_2 = 0, 1, 2, \ldots \}
 	\nonumber
 	\end{eqnarray}
be an orthonormal basis of ${\cal F}_{\kappa}$ with
 	\begin{eqnarray}
\langle n_1 , n_2 \vert n_1' , n_2' \rangle = \delta_{n_1,n_1'} \delta_{n_2,n_2'}. 
 	\nonumber
 	\end{eqnarray}
Number operators $N_1$ and $N_2$ are supposed to be diagonal in this basis, i.e.,  
		\begin{eqnarray}
N_i \vert n_1, n_2 \rangle = n_i \vert n_1, n_2 \rangle, \quad i=1,2
		\label{actionN}
		\end{eqnarray}
while the action of the creation and annihilation operators $a_1^{\pm}$ and $a_2^{\pm}$ is defined
by
		\begin{eqnarray}
a_1^+ \vert n_1, n_2 \rangle = \sqrt{F_1(n_1+1,n_2)} 
e^{{-i[H(n_1+1,n_2)- H(n_1, n_2)] \varphi }} \vert n_1+1, n_2 \rangle,
  	\label{action1+}
  	\end{eqnarray}
\begin{eqnarray}
a_1^- \vert n_1, n_2\rangle = \sqrt{F_1(n_1,n_2)} 
e^{{+i[H(n_1,n_2)- H(n_1-1, n_2)] \varphi }}\vert n_1-1,n_2\rangle, \quad a_1^- \vert 0 , n_2 \rangle = 0 
		\label{action1-}
		\end{eqnarray}
and 
		\begin{eqnarray}
a_2^+ \vert n_1, n_2 \rangle = \sqrt{F_2(n_1,n_2+1)} 
e^{{-i[H(n_1,n_2+1)- H(n_1, n_2)] \varphi }} \vert n_1,n_2+1 \rangle, 
		\label{action2+}
		\end{eqnarray}
		\begin{eqnarray}
a_2^- \vert n_1, n_2\rangle = \sqrt{F_2(n_1,n_2)} 
e^{{+i[H(n_1,n_2)- H(n_1, n_2-1)] \varphi }} \vert n_1,n_2-1\rangle, \quad a_2^- \vert n_1 , 0 \rangle = 0.
		\label{action2-}
		\end{eqnarray}
In Eqs.~(\ref{action1+})-(\ref{action2-}), $\varphi$ is an arbitrary real parameter and the 
positive valued functions $F_1 : \mathbb{N}^2 \to \mathbb{R}_+$, 
                          $F_2 : \mathbb{N}^2 \to \mathbb{R}_+$ and 
                          $H   : \mathbb{N}^2 \to \mathbb{R}_+$ are such that 
			\begin{eqnarray}
 H = F_1 + F_2.
			\nonumber 
			\end{eqnarray}
It is a simple matter of calculation to check that (\ref{actionN})-(\ref{action2-}) generate a representation 
of the ${\cal A}_{\kappa}(2)$ algebra defined by (\ref{commutation1})-(\ref{commutation3}) provided that 
$F_1(n_1 , n_2)$ and $F_2(n_1 , n_2)$ satisfy the recurrence relations 
		\begin{eqnarray}
&& F_1(n_1+1, n_2) - F_1(n_1, n_2) = 1 + \kappa ( 2n_1 + n_2), \quad F_1(0,n_2) = 0
		\label{recu1} \\
&& F_2(n_1, n_2+1) - F_2(n_1, n_2) = 1 + \kappa ( 2n_2 + n_1), \quad F_2(n_1,0) = 0.
		\label{recu2}
		\end{eqnarray}
The solutions of Eqs.~(\ref{recu1}) and (\ref{recu2}) are  
		\begin{eqnarray}
    F_i(n_1, n_2) = n_i [ 1 + \kappa (n_1+n_2-1)], \quad i=1,2.
    \label{Fi de n1,n2}
		\end{eqnarray}
To ensure that the structure functions $F_1$ and $F_2$ be positive definite, we must have 
		\begin{eqnarray}
1 + \kappa (n_1+n_2-1) > 0, \quad n_1 + n_2 > 0, 
		\label{condition}
		\end{eqnarray} 
a condition to be discussed according to the sign of $\kappa$. In the representation of ${\cal A}_{\kappa}(2)$ 
defined by Eqs.~(\ref{actionN})-(\ref{condition}), creation (annihilation) operators $a^+_i$ ($a^-_i$) and 
number operators $N_i$ satisfy the Hermitian conjugation relations
			\begin{eqnarray}
\left( a_i^- \right)^{\dagger} = a_i^+, \quad 
\left( N_i   \right)^{\dagger} = N_i, \quad i = 1, 2
 			\nonumber
			\end{eqnarray} 
as for the two-dimensional oscillator. 

The $d$ dimension of the representation space ${\cal F}_{\kappa}$ can be deduced from condition 
(\ref{condition}). Two cases need to be considered according to as $\kappa \geq 0$ or $\kappa < 0$.
\begin{itemize}
	\item For $\kappa \geq 0$, Eq.~(\ref{condition}) is trivially satisfied so that the $d$ dimension 
	of ${\cal F}_{\kappa}$ is infinite. This is well known in the case $\kappa = 0$ which corresponds 
	to a two-dimensional isotropic harmonic oscillator. For $\kappa > 0$, the representation corresponds 
	to the symmetric discrete (infinite-dimensional) irreducible representation of the $SU_{2,1}$ group. 
	\item For $\kappa < 0$, there exists a finite number of states satisfying condition 
	(\ref{condition}). Indeed, we have 
		\begin{eqnarray}
n_1 + n_2 = 0, 1, \ldots, E(-\frac{1}{\kappa}), 
		\nonumber 
		\end{eqnarray}
 where $E(x)$ denotes the integer part of $x$. In the following, we shall take $-1/\kappa$ 
 integer when $\kappa < 0$. Consequently, for $\kappa < 0$ the $d$ dimension of the 
 finite-dimensional space ${\cal F}_{\kappa}$ is 
		\begin{eqnarray}
d = \frac{1}{2}(k+1)(k+2), \quad k = -\frac{1}{\kappa} \in \mathbb{N}^*.
		\label{dimensiond}
		\end{eqnarray}
We know that the dimension $d(\lambda, \mu)$ of the irreducible representation $(\lambda, \mu)$
of $SU_3$ is given by 
		\begin{eqnarray}
d(\lambda, \mu) = \frac{1}{2}(\lambda + 1)(\mu + 1)(\lambda + \mu + 2 ), 
\quad \lambda \in \mathbb{N}, \quad \mu \in \mathbb{N}.
		\nonumber
		\end{eqnarray} 
Therefore, the finite-dimensional representation of ${\cal A}_{\kappa}(2)$ defined by (\ref{actionN})-(\ref{condition}) 
with $-1/\kappa = k \in \mathbb{N}^*$ corresponds to the irreducible representation $(0, k)$ or its adjoint $(k, 0)$ of 
$SU_3$. 
\end{itemize}

\subsection{Generalized oscillator Hamiltonian}

 Since the ${\cal A}_{\kappa}(2)$ algebra can be viewed
 as an extension of the two-dimensional oscillator algebra, it is natural to consider 
 the $a_1^+ a_1^- + a_2^+ a_2^-$ operator as an Hamiltonian associated with 
 ${\cal A}_{\kappa}(2)$. The action of this operator on the space ${\cal F}_{\kappa}$ is given 
 by
 			\begin{eqnarray}
 (a_1^+ a_1^- + a_2^+ a_2^-) \vert n_1,n_2 \rangle &=& 
 [F_1(n_1, n_2) + F_2(n_1, n_2)] \vert n_1,n_2 \rangle \nonumber \\ &=& 
 (n_1 + n_2) [1 + \kappa (n_1 + n_2 - 1)] \vert n_1,n_2 \rangle 
 			\nonumber 
 			\end{eqnarray}
 or 
 \begin{eqnarray}
 (a_1^+ a_1^- + a_2^+ a_2^-) \vert n_1,n_2 \rangle = H(n_1, n_2) \vert n_1,n_2 \rangle.
 \nonumber 
 \end{eqnarray}
 Thus, the $a_1^+ a_1^- + a_2^+ a_2^-$ Hamiltonian can be written
 \begin{eqnarray}
 a_1^+ a_1^- + a_2^+ a_2^- = H,
 \nonumber 
 \end{eqnarray}
 with 
 \begin{eqnarray}
 H \equiv H(N_1, N_2) = (N_1 + N_2) [1 + \kappa (N_1 + N_2 - 1)]
 \nonumber 
 \end{eqnarray}
 modulo its action on ${\cal F}_{\kappa}$. 
 
 The $H$ Hamiltonian is clearly a nonlinear extension of the Hamiltonian 
 for the two-dimensional isotropic harmonic oscillator. The eigenvalues 
 			\begin{eqnarray}
 \lambda_n = n [1 + \kappa (n - 1)], \quad n = n_1 + n_2, \quad n_1 \in \mathbb{N}, \quad n_2 \in \mathbb{N}
 			\nonumber
 			\end{eqnarray}
 of $H$ can be reduced for $\kappa = 0$ to the energies $n$ of the two-dimensional oscillator (up to additive 
 and multiplicative constants). For $\kappa \not= 0$, the degeneracy of the $\lambda_n$ level 
 is $n+1$ and coincides with the degeneracy of the $n$ level corresponding to $\kappa = 0$.   

\section{PHASE OPERATORS AND PHASE STATES FOR ${\cal A}_{\kappa}(2)$ WITH $\kappa < 0$}

\subsection{Phase operators in finite dimension}

\subsubsection{The $E_{1d}$ and $E_{2d}$ phase operators}

For $\kappa < 0$ the finite-dimensional space
${\cal F}_{\kappa}$ is spanned by the basis 
		\begin{eqnarray}
\{ \vert n_1 , n_2 \rangle : n_1, n_2 \ {\rm ranging} \ | \ n_1 + n_2 \leq k \}. 
		\nonumber
		\end{eqnarray}
This space can be partitioned as 
\begin{eqnarray}
{\cal F}_{\kappa} = \bigoplus_{l = 0}^{k}{\cal A}_{\kappa , l}, 
\nonumber 
\end{eqnarray}
where ${\cal A}_{\kappa , l}$ is spanned by 
\begin{eqnarray}
\{ \vert n , l \rangle : n = 0, 1, \ldots, k-l \}.
\nonumber 
\end{eqnarray}
We have 
		\begin{eqnarray}
{\rm dim \,} {\cal A}_{\kappa , l}  = k-l+1
		\nonumber
		\end{eqnarray}
so that (\ref{action1+})-(\ref{action1-}) must be completed by
		\begin{eqnarray}
a_1^+ \vert k-l , l \rangle = 0,
		\nonumber
		\end{eqnarray}
which can be deduced from the calculation of $\langle k-l , l \vert a_1^-a_1^+ \vert k-l , l \rangle$. The 
operators $a_1^+$ and $a_1^-$ leave each subspace ${\cal A}_{\kappa , l}$ invariant. Then, it is convenient 
to write
 		\begin{eqnarray}
 a_1^{\pm} = \sum_{l = 0}^{k} a_1^{\pm}(l), 
 		\nonumber
 		\end{eqnarray}
with the actions 
		\begin{eqnarray}
&& a_1^+(l) \vert n, l' \rangle = \delta_{l,l'} 
\sqrt{F_1(n+1,l)} e^{{-i[H(n+1,l)- H(n, l)] \varphi }} \vert n+1,l \rangle, 
		\nonumber \\
&& a_1^+(l) \vert k-l',l' \rangle = 0, 
		\nonumber \\
&& a_1^-(l) \vert n, l' \rangle = \delta_{l,l'} 
\sqrt{F_1(n,l)}   e^{{+i[H(n,l)- H(n-1, l)] \varphi }} \vert n-1,l \rangle, 
		\nonumber \\
&& a_1^-(l) \vert 0,l' \rangle = 0, 
		\nonumber 
		\end{eqnarray}
which show that $a_1^+(l)$ and $a_1^-(l)$ leave ${\cal A}_{\kappa , l}$ invariant. 

Let us now define the $E_{1d}$ operator by 
\begin{eqnarray}
E_{1d} \vert n_1 , n_2 \rangle = e^{i [H(n_1,n_2) - H(n_1-1,n_2)] \varphi} \vert n_1-1 , n_2 \rangle, \quad
0 \leq n_1 + n_2 \leq k, \quad n_1 \not= 0
\nonumber 
\end{eqnarray}
and 
\begin{eqnarray}
E_{1d} \vert 0 , n_2 \rangle = e^{i [H(0,n_2) - H(k-n_2,n_2)] \varphi} \vert k - n_2 , n_2 \rangle, \quad 
0 \leq n_2 \leq k, \quad n_1 = 0.
\nonumber 
\end{eqnarray}
Thus, it is possible to write 
		\begin{eqnarray}
a_1^- = E_{1d} \sqrt{F_1(N_1,N_2)} \Leftrightarrow  
a_1^+ =     \sqrt{F_1(N_1,N_2)} (E_{1d})^{\dagger}.
		\label{decomp pol 59}
		\end{eqnarray}
The $E_{1d}$ operator can be developed as 
\begin{eqnarray}
E_{1d} = \sum_{l = 0}^{k} E_{1d}(l), 
\nonumber 
\end{eqnarray}
with 
\begin{eqnarray}
E_{1d}(l) \vert n , l' \rangle &=& \delta_{l,l'} e^{i[H(n,l) - H(n-1,l)] \varphi} \vert n - 1, l  \rangle, \quad n \neq 0,
\label{actionE1} \\
E_{1d}(l) \vert 0 , l' \rangle &=& \delta_{l,l'} e^{i[H(0,l) - H(k-l,l)] \varphi} \vert k - l, l  \rangle, \quad n = 0.
\label{actionE1suite}
\end{eqnarray}
Operator $E_{1d}(l)$ leaves ${\cal A}_{\kappa , l}$ invariant and satisfies 
\begin{eqnarray}
E_{1d}(l) (E_{1d}(l'))^{\dagger} = (E_{1d}(l'))^{\dagger} E_{1d}(l) = 
\delta_{l,l'} \sum_{n = 0}^{k-l} \vert n , l \rangle \langle n , l \vert. 
\nonumber 
\end{eqnarray}
Consequently, we obtain
\begin{eqnarray}
E_{1d} (E_{1d})^{\dagger} = (E_{1d})^{\dagger} E_{1d} = \sum_{l = 0}^{k} (E_{1d}(l))^{\dagger} E_{1d}(l) =
 \sum_{l = 0}^{k} \sum_{n = 0}^{k-l} \vert n , l \rangle \langle n , l \vert = I, 
\nonumber 
\end{eqnarray}
which shows that $E_{1d}$ is unitary. Therefore, Eq.~(\ref{decomp pol 59}) constitutes a polar decomposition of 
$a_1^-$ and $a_1^+$. 

Similar developments can be obtained for $a_2^-$ and $a_2^+$. We 
limit ourselves to the main results concerning the decomposition 
\begin{eqnarray}
a_2^-  = E_{2d} \sqrt{F_2(N_1,N_2)} \Leftrightarrow  
a_2^+  =     \sqrt{F_2(N_1,N_2)} (E_{2d})^{\dagger}.
\nonumber 
\end{eqnarray}
In connection with this decomposition, we use the partition 
\begin{eqnarray}
{\cal F}_{\kappa} = \bigoplus_{l = 0}^{k} {\cal B}_{\kappa , l}, 
\nonumber 
\end{eqnarray}
where the ${\cal B}_{\kappa , l}$ subspace, of dimension $k-l+1$, is spanned by the basis 
\begin{eqnarray}
\{ \vert l , n \rangle : n = 0, 1, \ldots, k-l \}.
\nonumber 
\end{eqnarray}
We can write 
\begin{eqnarray}
E_{2d} = \sum_{l = 0}^{k} E_{2d}(l), 
\nonumber 
\end{eqnarray}
where the $E_{2d}(l)$ operator satisfies 
\begin{eqnarray}
E_{2d}(l) \vert l' , n \rangle &=& \delta_{l,l'} e^{i[H(l , n) - H(l, n-1)] \varphi} \vert l , n - 1 \rangle, \quad 
n \not = 0, \label{16prime} \\ 
E_{2d}(l) \vert l' , 0 \rangle &=& \delta_{l,l'} e^{i[H(l , 0) - H(l, k-l)] \varphi} \vert l , k - l \rangle, \quad 
n = 0, 
\label{17prime} 
\end{eqnarray}
and 
\begin{eqnarray}
E_{2d}(l) (E_{2d}(l'))^{\dagger} = (E_{2d}(l'))^{\dagger} E_{2d}(l) = 
\delta_{l,l'} \sum_{n = 0}^{k-l} \vert l , n \rangle \langle l , n \vert.
\nonumber 
\end{eqnarray}
This yields
		\begin{eqnarray}
E_{2d} (E_{2d})^{\dagger} = (E_{2d})^{\dagger} E_{2d} = \sum_{l = 0}^{k} (E_{2d}(l))^{\dagger} E_{2d}(l) = I
		\nonumber
		\end{eqnarray} 
and the operator $E_{2d}$, like $E_{1d}$, is unitary. 

\subsubsection{The $E_{3d}$ phase operator}

Let us go back to the pair $(a_3^+, a_3^-)$ of operators defined by (\ref{lesdeuxa3})
in terms of the pairs $(a_1^+, a_1^-)$ and $(a_2^+, a_2^-)$. The action of $a_3^+$ and 
$a_3^-$ on ${\cal F}_{\kappa}$ follows from (\ref{action1+})-(\ref{action2-}). We get 
		\begin{eqnarray}
a_3^+ \vert n_1, n_2 \rangle &=& - \kappa \sqrt{n_1(n_2 + 1)} \vert n_1-1 , n_2+1 \rangle, \nonumber \\
a_3^- \vert n_1, n_2 \rangle &=& - \kappa \sqrt{(n_1 + 1)n_2} \vert n_1+1 , n_2-1 \rangle. \nonumber
		\end{eqnarray}
From Eqs.~(\ref{commutation3}) and (\ref{lesdeuxa3}), it is clear that the two pairs ($a_1^+,a_1^-$) and 
($a_2^+,a_2^-$) commute when $\kappa = 0$. We thus recover that the ${\cal A}_0(2)$ algebra corresponds 
to a two-dimensional harmonic oscillator.
		
Here, it is appropriate to use the partition 
\begin{eqnarray}
{\cal F}_{\kappa} = \bigoplus_{l=0}^{k} {\cal C}_{\kappa , l}, 
\label{partitionC}
\end{eqnarray}
where the subspace ${\cal C}_{\kappa , l}$, of dimension $l+1$ (but not $k-l+1$ as 
for ${\cal A}_{\kappa , l}$ and ${\cal B}_{\kappa , l}$), spanned by the basis 
\begin{eqnarray}
\{ \vert l-n , n \rangle : n = 0, 1, \ldots, l \}
\nonumber 
\end{eqnarray}
is left invariant by $a_3^+$ and $a_3^-$. Following the same line of reasoning as for $E_{1d}$ and 
$E_{2d}$, we can associate an operator $E_{3d}$ with the ladder operators $a_3^+$ and $a_3^-$. We take 
operator $E_{3d}$ associated with the partition (\ref{partitionC}) such that
\begin{eqnarray}
a_3^-  = E_{3d} \sqrt{F_3(N_1,N_2)} \Leftrightarrow a_3^+ =  \sqrt{F_3(N_1,N_2)} (E_{3d})^{\dagger}, 
\nonumber 
\end{eqnarray}
where 
\begin{eqnarray}
\sqrt{F_3(N_1,N_2)} = - \kappa \sqrt{(N_1 + 1)N_2}.
\nonumber 
\end{eqnarray}
The $E_{3d}$ operator reads
\begin{eqnarray}
E_{3d} = \sum_{l=0}^{k} E_{3d}(l), 
\nonumber 
\end{eqnarray}
where $E_{3d}(l)$ can be taken to satisfy 
\begin{eqnarray}
E_{3d}(l) \vert l' - n , n \rangle = \delta_{l,l'} \vert l - n + 1 , n-1 \rangle, \quad n \neq 0,
\nonumber 
\end{eqnarray}
\begin{eqnarray}
E_{3d}(l) \vert l' , 0 \rangle = \delta_{l,l'} \vert 0 , l \rangle, \quad n = 0.
\nonumber 
\end{eqnarray}
Finally, we have 
		\begin{eqnarray}
E_{3d}(l) (E_{3d}(l'))^{\dagger} = (E_{3d}(l'))^{\dagger} E_{3d}(l) = 
\delta_{l,l'} \sum_{n = 0}^{l} \vert l-n , n \rangle \langle l-n , n \vert. 
		\nonumber
		\end{eqnarray}
As a consequence, we obtain
		\begin{eqnarray}
E_{3d} (E_{3d})^{\dagger} = (E_{3d})^{\dagger} E_{3d} = \sum_{l = 0}^{k} (E_{3d}(l))^{\dagger} E_{3d}(l) = I, 
		\nonumber
		\end{eqnarray}
a result that reflects the unitarity property of $E_{3d}$.

\subsubsection{The $E_{d}$ phase operator}

Operators $E_{1d}(l)$, $E_{2d}(l)$ and $E_{3d}(l)$, defined for $\kappa <0$ as components of the 
operators $E_{1d}$, $E_{2d}$ and $E_{3d}$, leave invariant the sets ${\cal A}_{\kappa,l}$, 
${\cal B}_{\kappa,l}$ and ${\cal C}_{\kappa,l}$, respectively. Therefore, operators 
$E_{1d}$, $E_{2d}$ and $E_{3d}$ do not connect all elements of ${\cal F}_{\kappa}$, i.e., a 
given element of ${\cal F}_{\kappa}$ cannot be obtained from repeated applications of 
$E_{1d}$, $E_{2d}$ and $E_{3d}$ on an arbitrary element of ${\cal F}_{\kappa}$. 

We now define a new operator $E_{d}$ which can connect (by means of repeated applications) 
any couple of elements in the $d$-dimensional space ${\cal F}_{\kappa}$ corresponding to 
$\kappa <0$. Let this global operator be defined via the action
			\begin{eqnarray}
E_{d} \vert n , l \rangle =  e^{i [H(n,l)- H(n-1, l)] \varphi} \vert n-1 , l \rangle, \quad 
n = 1, 2, \ldots, k-l, \quad  l = 0, 1, \ldots, k
			\label{Ed1}
			\end{eqnarray}
and the boundary actions
			\begin{eqnarray}
&& E_{d} \vert 0 , l \rangle =  e^{i [H(0,l)- H(k-l+1, l-1)] \varphi} \vert k-l+1, l-1 \rangle, 
\quad l = 1, 2, \ldots, k
			\label{Ed2} \\
&& E_{d} \vert 0 , 0 \rangle =  e^{i [H(0,0)- H(0, k)] \varphi} \vert 0, k \rangle. 
		\label{Ed3}
		\end{eqnarray}
The $E_{d}$ operator is obviously unitary. 

By making the identification 
			\begin{eqnarray}
\Phi_{\frac{1}{2}l(2k - l + 3) + n} \equiv \vert n , l \rangle, \quad n = 0, 1, \ldots, k-l, \quad l = 0, 1, \ldots, k, 
			\nonumber
			\end{eqnarray}
the set
			\begin{eqnarray}
\{ \Phi_{j} : j = 0, 1, \ldots, d-1 \}
			\nonumber
			\end{eqnarray}
constitutes a basis for the $d$-dimensional Fock space ${\cal F}_{\kappa}$. Then, 
the various sets ${\cal A}_{\kappa,l}$ can be rewritten as 
			\begin{eqnarray}
{\cal A}_{\kappa,0} &:& \{ \Phi_{0}, \Phi_{1}, \ldots, \Phi_{k-1}, \Phi_{k} \} \nonumber \\
{\cal A}_{\kappa,1} &:& \{ \Phi_{k+1}, \Phi_{k+2}, \ldots, \Phi_{2k} \} \nonumber \\
& \vdots & \nonumber \\
{\cal A}_{\kappa,k} &:& \{ \Phi_{d-1} \}. 
			\nonumber
			\end{eqnarray}
Repeated applications of $E_{d}$ on the vectors $\Phi_{j}$ with $j = 0, 1, \ldots, d-1$ can 
be summarized by the following cyclic sequence  
		\begin{eqnarray}
E_{d} : \Phi_{d-1} \mapsto \Phi_{d-2} \mapsto \ldots \mapsto \Phi_{1} \mapsto \Phi_{0} \mapsto \Phi_{d-1} \mapsto {\rm etc.}
		\nonumber
		\end{eqnarray}
The $E_{d}$ operator thus makes it possible to move inside each ${\cal A}_{\kappa,l}$ 
set and to connect the various sets according to the sequence 
		\begin{eqnarray}
		E_{d} : {\cal A}_{\kappa,k} \to {\cal A}_{\kappa,k-1} \to \ldots \to {\cal A}_{\kappa,0} \to {\cal A}_{\kappa,k} \to {\rm etc.}
		\nonumber
		\end{eqnarray}
Similar results hold for the partitions of ${\cal F}_{\kappa}$ in ${\cal B}_{\kappa,l}$ or ${\cal C}_{\kappa,l}$ subsets.

\subsection{Phase states in finite dimension}

\subsubsection{Phase states for $E_{1d}(l)$ and $E_{2d}(l)$}

We first derive the eigenstates of $E_{1d}(l)$. For 
this purpose, let us consider the eigenvalue equation
\begin{eqnarray}
E_{1d}(l) \vert z_l \rangle = z_l \vert z_l \rangle, \quad 
       \vert z_l \rangle = \sum_{n = 0}^{k-l} a_{n} z_l^n \vert n , l \rangle, \quad z_l \in \mathbb{C}.
\nonumber 
\end{eqnarray}
Using definition (\ref{actionE1})-(\ref{actionE1suite}), we obtain the following
recurrence relation for the coefficients $a_{n}$
\begin{eqnarray}
a_{n} = e^{-i [H(n , l) - H(n-1 , l)] \varphi} a_{n-1}, \quad n = 1, 2, \ldots, k-l
\nonumber 
\end{eqnarray}
with 
\begin{eqnarray}
a_{0} = e^{-i [H(0,l) - H(k-l,l)] \varphi} a_{k-l}
\nonumber 
\end{eqnarray}
and the condition
\begin{eqnarray}
(z_l)^{k-l+1} = 1.
\nonumber 
\end{eqnarray}
Therefore, we get
		\begin{eqnarray}
a_{n} = e^{-i [H(n,l) - H(0,l)] \varphi} a_{0}, \quad n = 0, 1, \ldots, k-l
		\nonumber 
		\end{eqnarray}
and the complex variable $z_l$ is a root of unity given by
\begin{eqnarray}
z_l = q_l^{m}, \quad m = 0, 1, \ldots, k-l,
\nonumber 
\end{eqnarray}
where
\begin{eqnarray}
q_l = \exp \left( \frac{2 \pi i} {k-l+1} \right) 
\label{definition of q_l}
\end{eqnarray}
is reminiscent of the deformation parameter used in the theory of quantum groups. The 
$a_{0}$ constant can be obtained, up to a phase factor, from the normalization condition 
$\langle z_l \vert z_l \rangle = 1$. We take 
\begin{eqnarray}
a_{0} = \frac{1}{\sqrt {k-l+1}} e^{-i H(0,l) \varphi},  
\label{choixdephase} 
\end{eqnarray}
where the phase factor is chosen in order to ensure temporal stability of the 
$\vert z_l \rangle$ state. Finally, we arrive at the following normalized 
eigenstates of $E_{1d}(l)$
\begin{eqnarray}
\vert z_l \rangle \equiv 
\vert l , m , \varphi \rangle = \frac{1}{\sqrt {k-l+1}}
\sum_{n=0}^{k-l} e^{-i H(n,l) \varphi} q_l^{mn} \vert n , l \rangle. 
\label{coherentstatemvarphi}
\end{eqnarray}
The $\vert l , m , \varphi \rangle$ states are labeled by the parameters $l \in \{ 0, 1, \ldots, k \}$, 
$m \in \mathbb{Z}/(k-l+1)\mathbb{Z}$ and $\varphi \in \mathbb{R}$. They satisfy
		\begin{eqnarray}
E_{1d}(l) \vert l , m , \varphi \rangle = e^{i\theta_{m}} \vert l , m , \varphi \rangle, \quad \theta_{m} = m \frac{2 \pi}{k-l+1}, 
\quad m = 0, 1, \ldots, k-l, 
		\label{ancienne92}
		\end{eqnarray}
which shows that $E_{1d}(l)$ is a phase operator.

The phase states $\vert l , m, \varphi \rangle$ have remarkable properties:
\begin{itemize}
	\item They are temporally stable with respect to the evolution operator 
	associated with the $H$ Hamiltonian. In other words, they satisfy
			\begin{eqnarray}
e^{-i H t} \vert l , m , \varphi \rangle = \vert l , m , \varphi + t \rangle
			\nonumber
			\end{eqnarray}
for any value of the real parameter $t$. 
	\item For fixed $\varphi$ and $l$, they satisfy the equiprobability relation
		\begin{eqnarray}
| \langle n , l \vert l , m , \varphi \rangle | =
\frac{1}{\sqrt{k-l+1}} 
		\nonumber
		\end{eqnarray}
and the property
			\begin{eqnarray}
\sum_{m = 0}^{k-l} \vert l , m , \varphi \rangle \langle l , m , \varphi \vert = 
\sum_{n=0}^{k-l} \vert n , l  \rangle \langle n , l \vert. 
			\nonumber
			\end{eqnarray}
	\item The overlap between two phase states $\vert l',  m' , \varphi' \rangle$ and 
$\vert l , m , \varphi \rangle$ reads
			\begin{eqnarray}
\langle l , m , \varphi \vert l' , m' , \varphi' \rangle = \delta_{l,l'} \frac{1}{k-l+1}
\sum_{n=0}^{k-l} q_l^{\rho(m-m', \varphi - \varphi', n)},
			\nonumber
			\end{eqnarray}
where
			\begin{eqnarray}
\rho(m-m', \varphi - \varphi', n) = - (m - m')n +
\frac{k-l+1}{2\pi} (\varphi - \varphi') H(n,l) 
			\nonumber
			\end{eqnarray}
with $q_l$ defined in (\ref{definition of q_l}). As a particular case, for fixed $\varphi$ we 
have the orthonormality relation
		\begin{eqnarray}
\langle l , m , \varphi \vert l', m' , \varphi \rangle = \delta_{l,l'} \delta_{m,m'}. 
    \nonumber
		\end{eqnarray}
However, not all temporally stable phase states are orthogonal.
\end{itemize}

Similar results can be derived for the $E_{2d}(l)$ operator by exchanging the roles played by 
$n$ and $l$. It is enough to mention that the $\vert z_l \rangle$ eigenstates of $E_{2d}(l)$ 
can be taken in the form 
			\begin{eqnarray}
\vert z_l \rangle \equiv 
\vert l , m , \varphi \rangle = \frac{1}{\sqrt {k-l+1}}
\sum_{n=0}^{k-l} e^{-i H(l,n) \varphi} q_l^{mn} \vert l , n \rangle
      \nonumber
			\end{eqnarray}
and present properties identical to those of the states in (\ref{coherentstatemvarphi}). 

\subsubsection{Phase states for $E_{3d}(l)$}

The eigenstates of the $E_{3d}(l)$ operator are given by 
\begin{eqnarray}
E_{3d}(l) \vert w_l \rangle = w_l \vert w_l \rangle, \quad 
          \vert w_l \rangle = \sum_{n = 0}^{l} c_{n} w_l^n \vert l-n , n \rangle, \quad w_l \in \mathbb{C}.
\nonumber 
\end{eqnarray}
The use of (\ref{16prime})-(\ref{17prime}) leads to the recurrence relation 
		\begin{eqnarray}
c_{n+1} = c_{n}, \quad n = 0, 1, \ldots, l-1
		\nonumber 
		\end{eqnarray}
with the condition
\begin{eqnarray}
c_{0} = c_{l} (w_l)^{l+1}.
\nonumber 
\end{eqnarray}
It follows that 
		\begin{eqnarray}
c_{n} = c_{0}, \quad n = 0, 1, \ldots, l
		\nonumber 
		\end{eqnarray}
and the $w_l$ eigenvalues satisfy 
		\begin{eqnarray}
(w_l)^{l+1} = 1.
		\nonumber 
	  \end{eqnarray}
Therefore, the admissible values for $w_l$ are 
		\begin{eqnarray}
w_l = \omega_l^{m}, \quad m = 0, 1, \ldots, l,
		\nonumber 
		\end{eqnarray}
with
		\begin{eqnarray}
\omega_l = \exp \left( \frac{2 \pi i} {l+1} \right).  
    \nonumber
		\end{eqnarray}
As a result, the normalized eigenstates of $E_{3d}(l)$ can be taken in the form 
			\begin{eqnarray}
\vert w_l \rangle \equiv 
\Vert l , m , \varphi \rangle \rangle = \frac{1}{\sqrt{l+1}} e^{-i H(0,l) \varphi}
\sum_{n=0}^{l} \omega_l^{mn} \vert l-n , n \rangle. 
    	\label{coherentstatemvarphiE3d}
			\end{eqnarray}
The $\Vert l , m , \varphi \rangle \rangle$ states depend on the parameters $l \in \{ 0, 1, \ldots, k \}$, 
$m \in \mathbb{Z}/(l+1)\mathbb{Z}$ and $\varphi \in \mathbb{R}$. They satisfy 
		\begin{eqnarray}
E_{3d}(l) \Vert l , m , \varphi \rangle \rangle = e^{i\theta_{m}} \Vert l , m , \varphi \rangle \rangle, 
\quad \theta_{m} = m \frac{2 \pi}{l+1},
    \label{ancienne92E3d}
		\end{eqnarray}
so that $E_{3d}(l)$ is a phase operator.

For fixed $l$, the set $\{ \Vert l , m , 0 \rangle \rangle : m = 0, 1, \ldots, l \}$, 
corresponding to $\varphi=0$, follows from the set $\{ \vert l-n , n \rangle : n = 0, 1, \ldots, l \}$
by making use of a (quantum) discrete Fourier transform.$^{\cite{Vourdas04}}$ Note that for $\varphi=0$, the 
$\Vert l , m , 0 \rangle \rangle$ phase states have the same form as the phase states for $SU_2$ derived by 
Vourdas.$^{\cite{Vourdas1990}}$ In the case where $\varphi \not= 0$, the 
$\Vert l , m , \varphi \rangle \rangle$ phase states for $E_{3d}(l)$ satisfy properties similar to those of 
the $\vert l , m , \varphi \rangle$ phase states for $E_{1d}(l)$ and for $E_{2d}(l)$ modulo the substitutions 
$(n , l) \to (l-n , n)$, $q_l \to \omega_l$ and $k-l \to l$.

\subsubsection{Phase states for $E_{d}$}

We are now in a position to derive the eigenstates of the $E_{d}$ 
operator. They are given by the following eigenvalue equation
		\begin{eqnarray}
E_{d} \vert \psi \rangle = \lambda \vert \psi \rangle, 
		\label{Eqlambda}
		\end{eqnarray}
where 
		\begin{eqnarray}
 \vert \psi \rangle = \sum_{l=0}^{k} \sum_{n=0}^{k-l} C_{n,l} \vert n,l \rangle.
		\label{vectlambda}
		\end{eqnarray}
Introducing (\ref{vectlambda}) into (\ref{Eqlambda}) and using the definition in (\ref{Ed1})-(\ref{Ed3}) 
of the $E_{d}$ operator, a straightforward but long calculation leads to following recurrence relations
		\begin{eqnarray}
&& C_{n+1,l} e^{i [H(n+1,l)- H(n, l)]   \varphi } = \lambda  C_{n,l} \label{rel-rec-1} \\
&& C_{0,l+1} e^{i [H(0,l+1)- H(k-l, l)] \varphi } = \lambda  C_{k-l,l}
		\label{rel-rec-2}
		\end{eqnarray}
for $l = 0, 1, \ldots, k-1$. For $l=k$, we have
			\begin{eqnarray}
C_{0,0} ~ e^{i [H(0,0)- H(0, k)] \varphi } = \lambda  C_{0,k}.
			\label{rel-rec-3}
			\end{eqnarray}
(Note that (\ref{rel-rec-2}) with $l=k$ yields (\ref{rel-rec-3}) if $C_{0,k+1}$ is identified to $C_{0,0}$.) From 
the recurrence relation (\ref{rel-rec-1}), it is  easy to get
			\begin{eqnarray}
 C_{n,l} = \lambda^n e^{{-i [H(n,l)- H(0, l)] \varphi }}  C_{0,l},
			\label{rel-3}
			\end{eqnarray}
which, for $n = k-l$, gives 
			\begin{eqnarray}
 C_{k-l,l} = \lambda^{k-l} e^{{-i [H(k-l,l)- H(0, l)] \varphi }}  C_{0,l}
			\label{soixante10} 
			\end{eqnarray}
in terms of $ C_{0,l}$. By introducing (\ref{soixante10}) into (\ref{rel-rec-2}), we obtain the recurrence relation
			\begin{eqnarray}
 C_{0,l+1} e^{{i [H(0,l+1)- H(0, l)] \varphi }} = \lambda^{k-l+1}   C_{0,l}
			\label{rel-4}
			\end{eqnarray}
that completely determines the $C_{0,l}$ coefficients and
subsequently the $C_{n,l}$ coefficients owing to (\ref{rel-3}). Indeed, the iteration of Eq.~(\ref{rel-4}) gives
			\begin{eqnarray}
 C_{0,l} = \lambda^{\frac{1}{2}l(2k - l + 3)} e^{-{i [H(0,l)- H(0, 0)] \varphi}} C_{0,0}.
			\label{soixante12} 
			\end{eqnarray} 
By combining (\ref{rel-3}) with (\ref{soixante12}), we finally obtain
		\begin{eqnarray}
 C_{n,l}= \lambda^{\frac{1}{2}l(2k - l + 3) + n}   e^{{-i H(n,l) \varphi }} C_{0,0}. 
		\label{avantavant73}
		\end{eqnarray} 
Note that for $l = k$ ($\Rightarrow n = 0$), Eq.~(\ref{avantavant73}) becomes
		\begin{eqnarray}
 C_{0,k}= \lambda^{\frac{1}{2}k(k + 3)} e^{{-i H(0,k) \varphi }} C_{0,0}. 
		\label{avant73}
		\end{eqnarray}
The introduction of (\ref{avant73}) in (\ref{rel-rec-3}) produces the condition
			\begin{eqnarray}
\lambda^d = 1.
			\nonumber 
			\end{eqnarray}
Consequently, the $\lambda$ eigenvalues are 
		\begin{eqnarray}
\lambda = \exp \left( \frac{2\pi i}{d} m \right), \quad m = 0, 1, \ldots, d-1.
		\nonumber
		\end{eqnarray}
Finally, the normalized eigenvectors of the $E_{d}$ operator read
		\begin{eqnarray} 
\vert \psi \rangle \equiv \vert m , \varphi \rangle = \frac{1}{\sqrt{d}}\sum_{l=0}^{k}
q^{\frac{1}{2}ml(2k-l+3)}\sum_{n=0}^{k-l} q^{mn} e^{-iH(n,l)\varphi} \vert n,l \rangle, 
		\label{m,phi}
		\end{eqnarray}
where
		\begin{eqnarray}
q = \exp \left( \frac{2\pi i}{d} \right).
		\label{definition of q}
		\end{eqnarray}
The $\vert m , \varphi \rangle$ states are labeled by the parameters $m
\in \mathbb{Z}/d\mathbb{Z}$ and $\varphi \in \mathbb{R}$. They satisfy
			\begin{eqnarray}
E_{d} \vert m , \varphi \rangle = e^{i\theta_m} \vert m , \varphi
\rangle, \quad \theta_m = m \frac{2 \pi}{d}, \quad m = 0, 1, \ldots, d-1.
			\nonumber 
			\end{eqnarray}
As a conclusion, $E_{d}$ is a unitary phase operator.

The $\vert m , \varphi \rangle$ phase states satisfy interesting properties: 
\begin{itemize}
	\item They are temporally stable under time evolution, i.e., 
		\begin{eqnarray}
e^{-i H t} \vert m , \varphi \rangle = \vert m , \varphi + t \rangle
    \nonumber
		\end{eqnarray}
for any value of the real parameter $t$.
	\item For fixed $\varphi$, they satisfy the relation
		\begin{eqnarray}
| \langle n , l \vert m , \varphi \rangle | = \frac{1}{\sqrt{d}}
		\nonumber
		\end{eqnarray}
and the closure property
			\begin{eqnarray}
\sum_{m = 0}^{d-1} \vert m , \varphi \rangle \langle m , \varphi
\vert = \sum_{l=0}^{k}\sum_{n=0}^{k-l} \vert n , l \rangle \langle n,l \vert = I.
			\nonumber			
			\end{eqnarray}					
	\item The overlap between two phase states $\vert m' , \varphi'
\rangle$ and $\vert m , \varphi \rangle$ reads
		\begin{eqnarray}
\langle m , \varphi \vert m' , \varphi' \rangle = \frac{1}{d}
\sum_{l=0}^{k}\sum_{n=0}^{k-l} q^{\tau(m'-m, \varphi - \varphi', n , l)}, 
		\nonumber
		\end{eqnarray}
where
		\begin{eqnarray}
\tau(m'-m, \varphi - \varphi', n, l) = (m' - m) \bigg[
\frac{1}{2}l(2k - l + 3) + n \bigg] + \frac{d}{2\pi} (\varphi - \varphi') H(n,l)
		\nonumber
		\end{eqnarray}
with $q$ defined in (\ref{definition of q}). As a particular case, we have the orthonormality relation
			\begin{eqnarray}
\langle m , \varphi \vert m' , \varphi \rangle = \delta_{m,m'}.
			\nonumber
			\end{eqnarray}
However, the temporally stable phase states are not all orthogonal.
\end{itemize}

\subsubsection{The $k=1$ particular case}

To close Section 3.2, we now establish a contact with the results of Klimov {\em et al.}$^{\cite{klimov1}}$ 
which correspond to $k=1$ (i.e., $\kappa = -1$). In this particular case, the ${\cal F}_{\kappa}$ 
Fock space is three-dimensional ($d=3$). It corresponds to the representation space of $SU_3$ 
relevant for ordinary quarks and antiquarks in particle physics and for qutrits in quantum 
information. For the purpose of comparison, we put   
			\begin{eqnarray}
\vert \phi_1 \rangle \equiv \vert 0,0 \rangle, \quad  
\vert \phi_2 \rangle \equiv \vert 1,0 \rangle, \quad
\vert \phi_3 \rangle \equiv \vert 0,1 \rangle.
			\nonumber
			\end{eqnarray}
Then, the operators $E_{13}$, $E_{23}$, $E_{33}$ and $E_{3}$ assume the form 
			\begin{eqnarray}
&& E_{13} = e^{ i \varphi} \vert \phi_1 \rangle \langle \phi_2 \vert + 
            e^{-i \varphi} \vert \phi_2 \rangle \langle \phi_1 \vert + 
                           \vert \phi_3 \rangle \langle \phi_3 \vert  \nonumber \\
&& E_{23} = e^{ i \varphi} \vert \phi_1 \rangle \langle \phi_3 \vert + 
            e^{-i \varphi} \vert \phi_3 \rangle \langle \phi_1 \vert + 
                           \vert \phi_2 \rangle \langle \phi_2 \vert  \nonumber \\
&& E_{33} =                \vert \phi_2 \rangle \langle \phi_3 \vert + 
                           \vert \phi_3 \rangle \langle \phi_2 \vert +
                           \vert \phi_1 \rangle \langle \phi_1 \vert  \nonumber \\
&& E_3    = e^{ i \varphi} \vert \phi_1 \rangle \langle \phi_2 \vert + 
                           \vert \phi_2 \rangle \langle \phi_3 \vert + 
            e^{-i \varphi} \vert \phi_3 \rangle \langle \phi_1 \vert. \nonumber
			\end{eqnarray}
Operators $E_{13}$, $E_{23}$ and $E_{33}$ have a form similar to that of the phase operators 
			\begin{eqnarray}
&& {\hat E}_{12} = \vert \phi_1 \rangle \langle \phi_2 \vert 
          - \vert \phi_2 \rangle \langle \phi_1 \vert  
          + \vert \phi_3 \rangle \langle \phi_3 \vert  \nonumber \\
&& {\hat E}_{13} = \vert \phi_1 \rangle \langle \phi_3 \vert 
          - \vert \phi_3 \rangle \langle \phi_1 \vert  
          + \vert \phi_2 \rangle \langle \phi_2 \vert  \nonumber \\
&& {\hat E}_{23} = \vert \phi_2 \rangle \langle \phi_3 \vert  
          - \vert \phi_3 \rangle \langle \phi_2 \vert
          + \vert \phi_1 \rangle \langle \phi_1 \vert  \nonumber 
			\end{eqnarray}
introduced in Ref.~{\cite{klimov1} in connection with qutrits. Although the $E_{13}$, $E_{23}$ and $E_{33}$ 
operators derived in the present work cannot be deduced from the ${\hat E}_{12}$, ${\hat E}_{13}$ and ${\hat E}_{23}$ operators
of Ref.~\cite{klimov1} by means of similarity transformations, the two sets of operators are equivalent in the 
sense that their action on the $\vert \phi_1 \rangle$, $\vert \phi_2 \rangle$ and $\vert \phi_3 \rangle$ 
vectors are identical up to phase factors. In addition, in the case where we do not take into account the 
spectator state ($\vert \phi_3 \rangle$, $\vert \phi_2 \rangle$ or $\vert \phi_1 \rangle$ for $E_{13}$, 
$E_{23}$ or $E_{33}$, respectively), our $SU_3$ phase operators are reduced to $SU_2$ phase operators 
which present the same periodicity condition (i.e., their square is the identity operator) as the $SU_2$ 
phase operators of Ref.~\cite{Vourdas1990}. In the $\varphi = 0$ case, our $SU_2$ phase 
states turn out to be identical to the phase states derived by Vourdas.$^{\cite{Vourdas1990}}$ Finally, note 
that the $E_{3}$ (and, more generally, $E_{d}$) operator is new; it has no equivalent in Ref.~\cite{klimov1}. 

\subsection{Vector phase states in finite dimension}

We have now the necessary tools for introducing vector phase states associated with the 
unitary phase operators $E_{1d}$, $E_{2d}$ and $E_{3d}$. We give below a construction 
similar to the one discussed in Ref.~\cite{twareque1}.

\subsubsection{Vector phase states for $E_{1d}$ and $E_{2d}$}

To define vector phase states, we introduce the $(k+1) \times (k+1)$-matrix
		\begin{eqnarray}
{\bf Z} = {\rm diag}(z_0 , z_1 , \ldots, z_k), \quad z_l = q_l^m
    \nonumber
		\end{eqnarray}
and the $(k+1) \times 1$-vector 
			\begin{eqnarray}
[ n , l ] =  \left(
\begin{array}{c}
  0 \\
  \vdots\\
  \vert n , l \rangle\\
  \vdots\\
  0\\
\end{array}
\right),
			\nonumber 
			\end{eqnarray}
where the $\vert n , l \rangle$ entry appears on the $l$-th line (with $l = 0, 1, \ldots, k$). Then, let us define 
			\begin{eqnarray}
[ l, m , \varphi ] = \frac{1}{\sqrt {k-l+1}}
    \sum_{n=0}^{k-l} e^{-i H(n,l) \varphi} {\bf Z}^{n} [n , l].  
			\label{vectorphasestates11}
			\end{eqnarray}
From Eq.~(\ref{coherentstatemvarphi}), we have
			\begin{eqnarray}
[ l, m , \varphi ] = \left(
\begin{array}{c}
  0 \\
  \vdots\\
  \vert l , m , \varphi \rangle\\
  \vdots\\
  0\\
\end{array}
\right),
			\label{vectorphasestates22}
			\end{eqnarray}
where $\vert l , m , \varphi \rangle$ occurs on the $l$-th line. 

We shall refer the states (\ref{vectorphasestates22}) to as vector phase states. In this matrix presentation, 
it is useful to associate the matrix 
\begin{eqnarray}
{\bf E_{1d}} = {\rm diag} \left( E_{1d}(0), E_{1d}(1), . . . ,E_{1d}(k) \right)
\nonumber 
\end{eqnarray}
with the unitary phase operator $E_{1d}$. It is easy to check 
that ${\bf E_{1d}}$ satisfies the matrix eigenvalue equation 
			\begin{eqnarray}
{\bf E_{1d}} [ l , m ,  \varphi ] = e^{i \theta_m} [ l , m ,  \varphi ], \quad \theta_m = m \frac{2 \pi}{k-l+1}
      \nonumber
			\end{eqnarray}
(cf.~Eq.~(\ref{ancienne92})). 

Other properties of vector phase states $[ l , m ,  \varphi ]$ can be deduced from those of phase states 
$\vert l , m ,  \varphi \rangle$. For instance, we obtain
\begin{itemize}
	\item The temporal stability condition
		\begin{eqnarray}
e^{-i H t} [ l , m ,  \varphi ] = [ l , m ,  \varphi + t] 
    \nonumber
		\end{eqnarray}
for $t$ real.
	\item The closure relation
			\begin{eqnarray}
\bigoplus_{l=0}^{k} \sum_{m=0}^{k-l} [ l , m ,  \varphi ] [ l , m ,  \varphi ]^{\dagger} = {\bf I_d}, 
			\nonumber
			\end{eqnarray}
where ${\bf I_d}$ is the unit matrix of dimension $d \times d$ with $d$ given by (\ref{dimensiond}).
\end{itemize}

Similar vector phase states can be obtained for $E_{2d}$ by permuting the $n$ and $l$ quantum numbers 
occurring in the derivation of the vector phase states for $E_{1d}$. 

\subsubsection{Vector phase states for $E_{3d}$}

Let us define the diagonal matrix of dimension $(k+1) \times (k+1)$
		\begin{eqnarray}
{\bf W} = {\rm diag}(w_0 , w_1 , \ldots, w_k), \quad w_l = \omega_l^m
    \nonumber
		\end{eqnarray}
and the column vector of dimension $(k+1) \times 1$
			\begin{eqnarray}
[[ n-l , n ]] =  \left(
\begin{array}{c}
  0 \\
  \vdots\\
  \vert l-n , n \rangle\\
  \vdots\\
  0\\
\end{array}
\right),
			\nonumber 
			\end{eqnarray}
where the $\vert l-n , n \rangle$ state occurs on the $l$-th line (with $l = 0, 1, \ldots, k$). By defining 
			\begin{eqnarray}
[[ l, m , \varphi ]] = \frac{1}{\sqrt {l+1}} e^{-i H(l,0) \varphi}
    \sum_{n=0}^{l} {\bf W}^{n} [[l-n , n]], 
			\nonumber
			\end{eqnarray}
we obtain
			\begin{eqnarray}
[[ l, m , \varphi ]] = \left(
\begin{array}{c}
  0 \\
  \vdots\\
  \Vert l , m , \varphi \rangle \rangle\\
  \vdots\\
  0\\
\end{array}
\right),
    	\label{vectorphasestates22E3d}
			\end{eqnarray}
where the $\Vert l , m , \varphi \rangle \rangle$ phase state appears on the $l$-th line. 

Equation (\ref{vectorphasestates22E3d}) defines vector phase states associated with the $E_{3d}$ phase operator. These 
states satisfy the eigenvalue equation  
			\begin{eqnarray}
			{\bf E_{3d}} [[ l , m ,  \varphi ]] = e^{i \theta_m} [[ l , m ,  \varphi ]], \quad \theta_m = m \frac{2 \pi}{l+1},
			\nonumber 
			\end{eqnarray}
where
			\begin{eqnarray}
{\bf E_{3d}} = {\rm diag} \left( E_{3d}(0), E_{3d}(1), . . . ,E_{3d}(k) \right). 
      \nonumber
			\end{eqnarray}
The $[[ l , m ,  \varphi ]]$ vector phase states satisfy properties which can be deduced from those of 
the $[  l , m ,  \varphi  ]$ vector phase states owing to simple correspondence rules. 

\section{PHASE OPERATORS AND PHASE STATES FOR ${\cal A}_{\kappa}(2)$ WITH $\kappa \geq 0$}

\subsection{Phase operators in infinite dimension}

In the case $\kappa \geq 0$, we can decompose the Jacobson operators $a_i^-$
and $a^+_i$ as
\begin{eqnarray}
a_i^- = E_{i\infty} \sqrt{F_i(N_1 , N_2)}, \quad 
a^+_i = \sqrt{F_i(N_1 , N_2)} \left( E_{i\infty} \right)^{\dagger}, \quad i = 1,2, 
\label{decompo cas infini}
\end{eqnarray}
where
\begin{eqnarray}
&& E_{1\infty} = \sum_{n_1=0}^{\infty}\sum_{n_2=0}^{\infty} e^{i
[H(n_1 + 1 , n_2)- H(n_1 , n_2)] \varphi } \vert n_1 , n_2 \rangle
\langle n_1+1 , n_2 \vert 
\label{E1infini} \\
&& E_{2\infty} = \sum_{n_1=0}^{\infty}\sum_{n_2=0}^{\infty} e^{i
[H(n_1 , n_2 +1)- H(n_1 , n_2)] \varphi } \vert n_1 , n_2 \rangle
\langle n_1, n_2 + 1\vert.
\label{E2infini}
\end{eqnarray}
The operators $E_{i\infty}$, $i = 1,2$, satisfy  
			\begin{eqnarray}
&& E_{1\infty}\left( E_{1\infty} \right)^{\dagger} = I, \quad 
\left( E_{1\infty} \right)^{\dagger} E_{1\infty} = 
I - \sum_{n_2=0}^{\infty} \vert 0 , n_2 \rangle\langle 0 , n_2\vert
			\label{pasunitaire1} \\
&& E_{2\infty}\left( E_{2\infty} \right)^{\dagger} = I, \quad 
\left( E_{2\infty} \right)^{\dagger} E_{2\infty} =
I - \sum_{n_1=0}^{\infty} \vert n_1 , 0 \rangle\langle n_1 , 0\vert. 
			\label{pasunitaire2} 
			\end{eqnarray}
Equations (\ref{pasunitaire1}) and (\ref{pasunitaire2}) show that $E_{i\infty}$, $i = 1,2$, are 
not unitary operators. 

In a similar way, operators $a_3^+$ and $a_3^-$ can be rewritten 
\begin{eqnarray}
a_3^- = - \kappa E_{3\infty} \sqrt{(N_1 + 1) N_2}, \quad 
a_3^+ = - \kappa \sqrt{(N_1 + 1) N_2} \left( E_{3\infty} \right)^{\dagger}, 
\nonumber
\end{eqnarray}
where 
		\begin{eqnarray}
E_{3\infty} = \sum_{n_1=0}^{\infty} 
              \sum_{n_2=0}^{\infty} \vert n_1 + 1 , n_2 \rangle \langle n_1 , n_2 + 1 \vert.
		\nonumber
		\end{eqnarray}
The $E_{3\infty}$ operator is not unitary since 
			\begin{eqnarray}
       E_{3\infty} \left( E_{3\infty} \right)^{\dagger} = I - \sum_{n_2=0}^{\infty} \vert   0 , n_2 \rangle \langle 0   , n_2 \vert, \quad 
\left( E_{3\infty} \right)^{\dagger}       E_{3\infty}  = I - \sum_{n_1=0}^{\infty} \vert n_1 , 0   \rangle \langle n_1 , 0   \vert, 
      \nonumber
			\end{eqnarray}
to be compared with (\ref{pasunitaire1}) and (\ref{pasunitaire2}).

The $E_{3\infty}$ operator is 
not independent of $E_{1\infty}$ and $E_{2\infty}$. Indeed, it can be expressed as 
			\begin{eqnarray}
      E_{3\infty} = \left( E_{1\infty} \right)^{\dagger} E_{2\infty},
      \label{pasunitaire3} 
			\end{eqnarray}
a relation of central importance for deriving its eigenvalues (see Section 4.2).

\subsection{Phase states in infinite dimension}

It is easy to show that operators $E_{1\infty}$ and $E_{2\infty}$ commute. Hence, 
that they can be simultaneously diagonalized. In this regard, let us consider 
the eigenvalue equations
\begin{eqnarray}
E_{1\infty} \vert z_1 , z_2 ) = z_1 \vert z_1 , z_2 ), \quad  
E_{2\infty} \vert z_1 , z_2 ) = z_2 \vert z_1 , z_2 ), \quad (z_1 , z_2) \in \mathbb{C}^2, 
\label{eqE1infini}
\end{eqnarray}
where
\begin{eqnarray}
\vert z_1 , z_2 ) = \sum_{n_1=0}^{\infty}
\sum_{n_2=0}^{\infty} D_{n_1 , n_2} \vert n_1, n_2 \rangle.
\nonumber 
\end{eqnarray}
By using the definitions of the nonunitary phase operators
(\ref{E1infini}) and (\ref{E2infini}), it is easy to check from the
eigenvalue equations (\ref{eqE1infini})  that the complex
coefficients $D_{n_1 , n_2}$ satisfy the following recurrence
relations
\begin{eqnarray}
&& D_{n_1 + 1 , n_2} e^{i H(n_1+1 , n_2) \varphi} = z_1 D_{n_1 , n_2} e^{iH(n_1 , n_2)\varphi} 
 \label{recurencekappa+1} \\
&& D_{n_1 , n_2 + 1} e^{i H(n_1 , n_2+1) \varphi} = z_2 D_{n_1 , n_2} e^{iH(n_1 , n_2)\varphi},  
\label{recurencekappa+2}
\end{eqnarray}
which lead to
\begin{eqnarray}
D_{n_1 , n_2}= e^{-i H(n_1 , n_2)  \varphi} z_1^{n_1}  z_2^{n_2}D_{0, 0}.
\nonumber 
\end{eqnarray}
It follows that the normalized common  eigenstates of the 
operators $E_{1\infty}$ and $E_{2\infty}$ are given by 
\begin{eqnarray}
\vert z_1 , z_2 ) = \sqrt{(1 - |z_1|^2)(1 - |z_2|^2)}
\sum_{n_1=0}^{\infty}\sum_{n_2=0}^{\infty} z_1^{n_1} z_2^{n_2}e^{- i
H(n_1 , n_2)\varphi} \vert n_1 , n_2 \rangle
\nonumber 
\end{eqnarray}
on the domain $\{ (z_1 , z_2)  \in {\mathbb{C}^2} : |z_1| < 1, |z_2| < 1 \}$. Following the method developed in
Refs.~\cite{vourdasLimit, voudasBM} for the Lie algebra $su_{1,1}$ and in Ref.~\cite{daoud-kibler} for the algebra 
${\cal A}_{\kappa}(1)$, we define the states 
\begin{eqnarray}
\vert \theta_1, \theta_2,\varphi ) = \lim_{z_1 \rightarrow
e^{i\theta_1}} \lim_{z_2 \rightarrow e^{i\theta_2}}
\frac{1}{\sqrt{(1 - |z_1|^2)(1 - |z_2|^2)}} \vert z_1 , z_2 ),
\nonumber 
\end{eqnarray}
where $\theta_1 , \theta_2 \in [-\pi , +\pi]$. We thus get 
\begin{eqnarray}
\vert \theta_1, \theta_2,\varphi ) =
\sum_{n_1=0}^{\infty}\sum_{n_2=0}^{\infty} e^{i n_1 \theta_1}e^{i
n_2 \theta_2} e^{- i H(n_1 , n_2) \varphi} \vert n_1 , n_2 \rangle.
\nonumber 
\end{eqnarray}
These states, defined on $S^1 \times S^1$, turn out to be phase
states since we have
\begin{eqnarray}
E_{1\infty} \vert \theta_1, \theta_2,\varphi ) = e^{i
\theta_1}  \vert \theta_1, \theta_2,\varphi ), \quad
E_{2\infty} \vert \theta_1, \theta_2,\varphi ) = e^{i
\theta_2}  \vert \theta_1, \theta_2,\varphi ).
\nonumber 
\end{eqnarray}
Hence, the operators $E_{i\infty}$, $i=1,2$, are (nonunitary) phase operators.

The main properties of the $\vert \theta_1 , \theta_2 , \varphi )$ states 
are the following. 
\begin{itemize}
	\item They are temporally stable in the sense that
		\begin{eqnarray}
e^{-i H t} \vert \theta_1 , \theta_2 , \varphi ) = \vert \theta_1 , \theta_2 , \varphi + t ), 
		\nonumber
		\end{eqnarray}
with $t$ real. 
	\item They are not normalized and not
orthogonal. However, for fixed $\varphi$, they satisfy the closure
relation
		\begin{eqnarray}
\frac{1}{(2\pi)^2} \int_{-\pi}^{+\pi} d\theta_1  \int_{-\pi}^{+\pi}
d\theta_2 \vert \theta_1 , \theta_2  , \varphi ) ( \theta_1 , \theta_2  , \varphi \vert = I.
		\nonumber
		\end{eqnarray}
\end{itemize}

In view of Eq.~(\ref{pasunitaire3}), we have 
			\begin{eqnarray}
       E_{3\infty} | \theta_1 , \theta_2, \varphi ) = e^{i(\theta_2 - \theta_1)} | \theta_1 , \theta_2, \varphi ),
      \nonumber
			\end{eqnarray}
so that the $| \theta_1 , \theta_2, \varphi )$ states are common eigenstates to $E_{1\infty}$, $E_{2\infty}$ and $E_{3\infty}$.

To close, a comparison is in order. For $\varphi = 0$, the $\vert \theta_1 , \theta_2 , 0 )$ states 
have the same form as the phase states derived in Ref.~\cite{bertola-Deguise} which present the closure 
property but are not temporally stable. 

\section{TRUNCATED GENERALIZED OSCILLATOR ALGEBRA}
			
For $\kappa \geq 0$ the ${\cal F}_{\kappa}$ Hilbert space associated with ${\cal A}_{\kappa}(2)$ 
is infinite-dimensional and it is thus impossible to define a unitary phase operator. On the other 
hand, for $\kappa < 0$ the ${\cal F}_{\kappa}$ space is finite-dimensional and there is no problem 
to define unitary phase operators. Therefore, for $\kappa \geq 0$ it is appropriate to truncate the 
${\cal F}_{\kappa}$ space in order to get a subspace ${\cal F}_{\kappa , \sigma}$ of dimension 
$(\sigma + 1)(\sigma + 2)/2$ with $\sigma$ playing the role of $k$. Then, it will be possible 
to define unitary phase operators and vector phase vectors for the ${\cal F}_{\kappa , \sigma}$ 
truncated space with $\kappa \geq 0$. To achieve this goal, we shall adapt the truncation procedure 
discussed in Ref.~\cite{pegg-barnett} for the $h_4$ Weyl-Heisenberg algebra and in Ref.~\cite{AKW,daoud-kibler} 
for the ${\cal A}_{\kappa}(1)$ algebra with $\kappa \geq 0$.  

The restriction of infinite-dimensional space ${\cal F}_{\kappa}$ ($\kappa \geq 0$) to 
finite-dimensional space ${\cal F}_{\kappa,\sigma}$ with basis 
			\begin{eqnarray}
      \{ |n_1 , n_2 \rangle : n_1, n_2 \ {\rm ranging} \ | \ n_1+n_2 \leq \sigma \}
      \nonumber
			\end{eqnarray}
can be done by means of the projection operator 
			\begin{eqnarray}
      \Pi_{\sigma} = \sum_{n_1 = 0}^{\sigma} \sum_{n_2 = 0}^{\sigma - n_1} \vert n_1 , n_2 \rangle \langle n_1 , n_2 \vert =
                     \sum_{n_2 = 0}^{\sigma} \sum_{n_1 = 0}^{\sigma - n_2} \vert n_1 , n_2 \rangle \langle n_1 , n_2 \vert.
      \nonumber
			\end{eqnarray}
Let us then define the four new ladder operators
			\begin{eqnarray}
      b_i^{\pm} = \Pi_{\sigma} a_i^{\pm} \Pi_{\sigma}, \quad i = 1, 2.
      \nonumber
			\end{eqnarray}
They can be rewritten as 
			\begin{eqnarray}
      && b_1^{+} = (b_1^{-})^{\dagger} = \sum_{n_2 = 0}^{\sigma-1} \sum_{n_1 = 0}^{\sigma - n_2-1}
			\sqrt{F_1(n_1+1,n_2)} e^{-i [H(n_1+1,n_2) - H(n_1, n_2)] \varphi}
			\vert n_1+1 , n_2 \rangle \langle n_1 , n_2 \vert \nonumber \\
      && b_2^{+} = (b_2^{-})^{\dagger} = \sum_{n_1 = 0}^{\sigma-1} \sum_{n_2 = 0}^{\sigma - n_1-1}
			\sqrt{F_2(n_1,n_2+1)} e^{-i [H(n_1,n_2+1)- H(n_1, n_2)] \varphi}
			\vert n_1 , n_2+1 \rangle \langle n_1 , n_2 \vert
      \nonumber
			\end{eqnarray}
A straightforward calculation shows that the action of $b_1^{\pm}$ on ${\cal F}_{\kappa}$ is given by 
		\begin{eqnarray}
		&& b_1^+ \vert n_1, n_2 \rangle = \sqrt{F_1(n_1+1,n_2)} 
		e^{-i[H(n_1+1,n_2)- H(n_1, n_2)] \varphi} \vert n_1+1, n_2 \rangle \nonumber \\ 
    && \qquad \qquad \qquad \qquad \qquad \qquad \qquad {\rm for} \quad n_1+n_2 = 0, 1, \ldots, \sigma-1 \nonumber \\
		&& b_1^+ \vert \sigma - n_2, n_2 \rangle = 0 \quad {\rm for} \quad n_2 = 0, 1, \ldots, \sigma \nonumber \\
		&& b_1^+ \vert n_1, n_2          \rangle = 0 \quad {\rm for} \quad n_1+n_2 = \sigma, \sigma+1, \sigma+2, \ldots \nonumber
  	\end{eqnarray}
and
		\begin{eqnarray}
		&& b_1^- \vert n_1, n_2 \rangle = \sqrt{F_1(n_1,n_2)} 
		e^{+i[H(n_1,n_2)- H(n_1 - 1, n_2)] \varphi} \vert n_1 - 1, n_2 \rangle \nonumber \\
    && \qquad \qquad \qquad \qquad \qquad {\rm for} \quad n_1 \not= 0 \quad {\rm and} \quad n_2 = 0, 1, \ldots, \sigma - 1 \nonumber \\ 
    && b_1^- \vert 0 , n_2  \rangle = 0 \quad {\rm for} \quad n_2 = 0, 1, \ldots, \sigma \nonumber \\
		&& b_1^- \vert n_1, n_2 \rangle = 0 \quad {\rm for} \quad n_1+n_2 = \sigma+1, \sigma+2, \sigma+3, \ldots. \nonumber
  	\end{eqnarray}
Similarly, we have 
		\begin{eqnarray}
		&& b_2^+ \vert n_1, n_2 \rangle = \sqrt{F_2(n_1,n_2+1)} 
		e^{-i[H(n_1,n_2+1)- H(n_1, n_2)] \varphi} \vert n_1, n_2+1 \rangle \nonumber \\ 
    && \qquad \qquad \qquad \qquad \qquad \qquad \qquad {\rm for} \quad n_1+n_2 = 0, 1, \ldots, \sigma-1 \nonumber \\ 
		&& b_2^+ \vert n_1 , \sigma - n_1 \rangle = 0 \quad {\rm for} \quad n_1 = 0, 1, \ldots, \sigma \nonumber \\
		&& b_2^+ \vert n_1, n_2           \rangle = 0 \quad {\rm for} \quad n_1+n_2 = \sigma, \sigma+1, \sigma+2, \ldots \nonumber
  	\end{eqnarray}
and
		\begin{eqnarray}
		&& b_2^- \vert n_1, n_2 \rangle = \sqrt{F_2(n_1,n_2)} 
		e^{+i[H(n_1,n_2)- H(n_1, n_2 - 1)] \varphi} \vert n_1 , n_2 - 1 \rangle \nonumber \\ 
    && \qquad \qquad \qquad \qquad \qquad {\rm for} \quad n_2 \not= 0 \quad {\rm and} \quad n_1 = 0, 1, \ldots, \sigma - 1 \nonumber \\ 
    && b_2^- \vert n_1 , 0  \rangle = 0 \quad {\rm for} \quad n_1 = 0, 1, \ldots, \sigma \nonumber \\
		&& b_2^- \vert n_1, n_2 \rangle = 0 \quad {\rm for} \quad n_1+n_2 = \sigma+1, \sigma+2, \sigma+3, \ldots. \nonumber
  	\end{eqnarray}
Therefore, the action of operators $b_i^{\pm}$ ($i=1,2$) on ${\cal F}_{\kappa,\sigma}$ with $\kappa \geq 0$ is 
similar to that of $a_i^{\pm}$ ($i=1,2$) on ${\cal F}_{\kappa}$ with $\kappa < 0$.

We may ask what is the algebra generated by operators $b_i^{\pm}$ and $N_i$ ($i=1,2$)? Indeed, the latter 
operators satisfy the following algebraic relations when acting on the ${\cal F}_{\kappa,\sigma}$ space
   		\begin{eqnarray}
    & & [b_1^- , b_1^+] = I + \kappa (2 N_1 + N_2) 
        - \sum_{l=0}^{\sigma} F_1(\sigma - l +1, l) \vert \sigma-l , l \rangle \langle \sigma-l , l \vert \nonumber \\
    & & [b_2^- , b_2^+] = I + \kappa (2 N_2 + N_1) 
        - \sum_{l=0}^{\sigma} F_2(l, \sigma - l +1) \vert l , \sigma-l \rangle \langle l , \sigma-l \vert \nonumber \\  		
   	& & [N_i , b_j^{\pm}] = {\pm} \delta_{i,j} b_i^{\pm}, \quad i,j = 1,2
			\nonumber \\
    & & [b_i^{\pm} , b_j^{\pm}] = 0, \quad [b_i^{\pm} , [b_i^{\pm} , b_j^{\mp}]] = 0, \quad i \neq j. 
			\nonumber
   		\end{eqnarray}
Operators $b_i^{\pm}$ and $N_i$ ($i=1,2$) acting on ${\cal F}_{\kappa,\sigma}$ generate an algebra, noted 
${\cal A}_{\kappa, \sigma}(2)$. The 
${\cal A}_{\kappa, \sigma}(2)$ algebra generalizes ${\cal A}_{\kappa, s}(1)$ which results from 
the truncation of the ${\cal A}_{\kappa}(1)$ algebra.$^{\cite{daoud-kibler}}$ By using the trick 
to pass from ${\cal A}_{\kappa}(2)$ to ${\cal A}_{\kappa}(1)$, see section 2.1, we get 
${\cal A}_{\kappa, s-1}(2) \to {\cal A}_{\kappa, s}(1)$. The ${\cal A}_{\kappa, s}(1)$ truncated algebra 
gives in turn the Pegg-Barnett truncated algebra$^{\cite{pegg-barnett}}$ when $\kappa \to 0$. 

As a conclusion, the action of $b_i^{\pm}$ ($i=1,2$) on the complement of ${\cal F}_{\kappa, \sigma}$ with respect to ${\cal F}_{\kappa}$
leads to the null vector while the action of these operators on the ${\cal F}_{\kappa, \sigma}$ space with $\kappa \geq 0$ is the same as 
the action of $a_i^{\pm}$ ($i=1,2$) on the ${\cal F}_{\kappa}$ space with $\kappa < 0$ modulo some evident changes of notations. It is 
thus possible to apply the procedure developed for ${\cal F}_{\kappa}$ space with $\kappa < 0$ in order to obtain unitary phase operators 
on ${\cal F}_{\kappa, \sigma}$ with $\kappa \geq 0$ and the corresponding vector phase states. The derivation of the vector phase states 
for the ${\cal A}_{\kappa, \sigma}(2)$ truncated algebra can be done simply by replacing $k$ by $\sigma$. In this respect, the 
$\sigma$ truncation index can be compared to the $k$ quenching index (or Chen index) used for characterizing the finite-dimensional 
representation $(0,k)$ or $(k,0)$ of $SU_3$.$^{\cite{Chenindex}}$ 

\section{APPLICATION TO MUTUALLY UNBIASED BASES} 

We now 
examine the possibility to produce specific bases, known as mutually unbiased bases (MUBs) 
in quantum information, for finite-dimensional Hilbert spaces from the phase states of 
$E_{1d}(l)$, $E_{2d}(l)$ and $E_{3d}(l)$. Let us recall that two distinct orthonormal bases 
      \begin{eqnarray}
\{ | a \alpha \rangle : \alpha = 0, 1, \ldots, N-1 \}
      \nonumber
      \end{eqnarray}
and
      \begin{eqnarray}
\{ | b \beta  \rangle : \beta  = 0, 1, \ldots, N-1 \}
      \nonumber
      \end{eqnarray}
of the $N$-dimensional Hilbert spaces $\mathbb{C}^{N}$ are said to be unbiased if and only if
      \begin{eqnarray}
\forall \alpha = 0, 1, \ldots, N-1, \ \
\forall \beta  = 0, 1, \ldots, N-1 \ : \ \vert \langle a \alpha | b \beta \rangle \vert = \frac{1}{\sqrt{N}}
      \nonumber
      \end{eqnarray}
(cf.~Refs.~\cite{ivanovic,Klimov05,Klimov06,woottersFields}). 

We begin with the $\vert l , m , \varphi \rangle$ phase states associated with the $E_{1d}(l)$ phase operator 
(see (\ref{coherentstatemvarphi}) and (\ref{ancienne92})). In Eq.~(\ref{coherentstatemvarphi}), $l$ can take the values 
$0, 1, \ldots, k$. Let us put $l=0$ and switch to the notations
		\begin{eqnarray}
k \equiv N-1, \quad m \equiv \alpha, \quad | n,0 \rangle \equiv | N-1-n \rangle
		\nonumber
		\end{eqnarray}
(with $\alpha, n = 0, 1, \ldots, N-1$) for easy comparison with some previous works. Then, Eq.~(\ref{coherentstatemvarphi}) 
becomes 
			\begin{eqnarray}
\vert 0 , \alpha , \varphi \rangle = \frac{1}{\sqrt {N}}
\sum_{n=0}^{N-1} \exp \left[ -\frac{i}{N-1} n(N-n) \varphi + \frac{2 \pi i}{N} n \alpha \right] \vert N-1-n \rangle. 
			\label{zeroalphaphi}
			\end{eqnarray}
For $\varphi=0$, Eq.~(\ref{zeroalphaphi}) describes a (quantum) discrete Fourier transform$^{\cite{Vourdas04}}$ that allows to pass 
from the set $\{ |N - 1 - n \rangle : n = 0, 1, \ldots, N-1 \}$ of cardinal $N$ to the set 
$\{ \vert 0 , \alpha , 0 \rangle : \alpha = 0, 1, \ldots, N-1 \}$ of cardinal $N$ too. In the special case where $\varphi$ 
is quantized as 
			\begin{eqnarray}
\varphi = - \pi \frac{N-1}{N} a,  \quad a = 0, 1, \ldots, N-1,  
			\label{phidiscrete}
			\end{eqnarray}
equation (\ref{zeroalphaphi}) leads to 
			\begin{eqnarray}
\vert 0 , \alpha , \varphi \rangle \equiv | a \alpha \rangle = \frac{1}{\sqrt {N}}
\sum_{n=0}^{N-1} q_0^{n(N-n) a/2 + n \alpha} \vert N-1-n \rangle,  
			\label{zeroalphaphiquantized}
			\end{eqnarray}
where 
			\begin{eqnarray}
q_0 = \exp \left( \frac{2 \pi i} {N} \right). 
      \nonumber 
			\end{eqnarray}
Equation (\ref{zeroalphaphiquantized}) with $a \not= 0$ corresponds to a (quantum) 
quadratic discrete Fourier trans\-form.$^{\cite{kibler3,kibler1,kibler2}}$ In this regard, note that the 
$| a \alpha \rangle$ state in (\ref{zeroalphaphiquantized}) can be identified with the $| a \alpha ; r \rangle$ 
state with $r=0$ discussed recently in the framework of the quadratic discrete Fourier transform.$^{\cite{InTech}}$ Following 
Ref.~\cite{InTech}, we consider the set 
			\begin{eqnarray}
B_N = \{ | N-1-n \rangle : n = 0, 1, 2, \ldots, N-1 \} = \{ | n \rangle : n = 0, 1, 2, \ldots, N-1 \}
			\nonumber
			\end{eqnarray}
as an orthonormal basis for the $N$-dimensional Hilbert space. This basis is called computational basis in 
quantum information. Then, the sets 
			\begin{eqnarray}
B_{0a} = \{ | a \alpha \rangle : \alpha = 0, 1, 2, \ldots, N-1 \}, \quad a = 0, 1, 2, \ldots, N-1 
			\nonumber
			\end{eqnarray}
constitute $N$ new orthonormal bases of the space. The $B_{0a}$ basis is a special case, corresponding to $r=0$, 
of the $B_{ra}$ bases derived in Ref.~\cite{InTech} from a polar decomposition of the $su_2$ Lie algebra. The overlap 
between two bases $B_{0a}$ and $B_{0b}$ is given by 
			\begin{eqnarray}
\langle a \alpha | b \beta \rangle = \frac{1}{N} 
\sum_{n=0}^{N-1} q_0^{n(N-n) (b-a)/2 + n (\beta - \alpha)},  
			\nonumber
			\end{eqnarray}
a relation which can be expressed in term of the generalized Gauss sum$^{\cite{les2Berndt}}$
			\begin{eqnarray}
S(u, v, w) = \sum_{k=0}^{|w|-1} e^{i \pi (uk^2 + vk) / w}.
			\nonumber
			\end{eqnarray}
In fact, we obtain 
			\begin{eqnarray}
\langle a \alpha | b \beta \rangle = \frac{1}{N} S(u, v, w),  
			\label{overlap en S}
			\end{eqnarray}
with 
			\begin{eqnarray}
u = a-b, \quad v = -(a-b)N - 2 (\alpha-\beta), \quad w = N.
			\nonumber
			\end{eqnarray}
In the case where $N$ is a prime integer, the calculation of $S(u, v, w)$ in (\ref{overlap en S}) yields 
			\begin{eqnarray}
\vert \langle a \alpha | b \beta \rangle \vert = \frac{1}{\sqrt{N}}, 
\quad a \not= b, \quad \alpha, \beta = 0, 1, \ldots, N-1, \quad N \ {\rm prime}.
			\label{MUB1}
			\end{eqnarray}
On the other hand, it is evident that 
			\begin{eqnarray}
\vert \langle n | a \alpha \rangle \vert = \frac{1}{\sqrt{N}}, \quad n, \alpha = 0, 1, \ldots, N-1
			\label{MUB2}
			\end{eqnarray}
holds for any strictly positive value of $N$. As a result, Eqs.~(\ref{MUB1}) and (\ref{MUB2}) shows that 
bases $B_N$ and $B_{0a}$ with $a = 0, 1, \ldots, N-1$ provide a complete set of $N+1$ MUBs when $N$ is a 
prime integer.  

A similar result can be derived by quantizing, according to (\ref{phidiscrete}), the $\varphi$ 
parameter occurring in the eigenstates of $E_{2d}(0)$. 

The form of the $E_{3d}(l)$ phase operator being different from those of $E_{1d}(l)$ and  $E_{2d}(l)$, we proceed 
in a different way for obtaining MUBs from the $\Vert l , m , \varphi \rangle \rangle$ eigenstates of $E_{3d}(l)$ 
(see (\ref{coherentstatemvarphiE3d}) and (\ref{ancienne92E3d})). We put $\varphi=0$ in (\ref{coherentstatemvarphiE3d}) 
and apply the $e^{- i F_3(N_1,N_2) \varphi}$ operator on the resultant state. This gives 
		\begin{eqnarray}
e^{- i F_3(N_1,N_2) \varphi} \Vert l , m , 0 \rangle \rangle = \frac{1}{\sqrt{l+1}}
\sum_{n=0}^{l} \exp \left[ -i \frac{1}{k^2} n(l+1-n) \varphi \right] \omega_l^{mn} \vert l-n , n \rangle. 
		\nonumber
		\end{eqnarray} 
For the sake of comparison, we introduce
		\begin{eqnarray}
l \equiv N-1, \quad m \equiv \alpha, \quad \omega_{N-1} \equiv \exp \left( \frac{2 \pi i}{N} \right), \quad | l-n,n \rangle \equiv | N-1-n \rangle
		\nonumber
		\end{eqnarray}
and we quantize $\varphi$ via
			\begin{eqnarray}
\varphi = - \pi \frac{k^2}{N} a,  \quad a = 0, 1, \ldots, N-1.  
			\nonumber
			\end{eqnarray}
Hence, the vector 
			\begin{eqnarray}
e^{- i F_3(N_1,N_2) \varphi} \Vert l , m , 0 \rangle \rangle \equiv | a \alpha \rangle
			\nonumber
			\end{eqnarray}
reads
			\begin{eqnarray}
\vert a \alpha \rangle = \frac{1}{\sqrt {N}}
\sum_{n=0}^{N-1} \omega_{N-1}^{n(N-n) a/2 + n \alpha} \vert N-1-n \rangle,  
			\label{aalphaE3d(l)}
			\end{eqnarray}
which bears the same form as (\ref{zeroalphaphiquantized}). Consequently for $N$ a prime 
integer, Eq.~(\ref{aalphaE3d(l)}) generates $N$ MUBs $B_{0a}$ with $a = 0, 1, \ldots, N-1$ which 
together with the computational basis $B_N$ form a complete set of $N+1$ MUBs.

\section{CONCLUDING REMARKS}

The main results of this work are the following. 

The $su_3$, $su_{2,1}$ and $h_4 \otimes h_4$ algebras 
can be described in an unified way via the introduction of the ${\cal A}_{\kappa}(2)$ algebra. A quantum system 
with a quadratic spectrum (for $\kappa \not= 0$) is associated with ${\cal A}_{\kappa}(2)$~; for $\kappa = 0$, 
this system coincides with the two-dimensional isotropic harmonic oscillator. 

In the case $\kappa < 0$, 
the unitary phase operators ($E_{1d}$, $E_{2d}$ and $E_{3d}$) defined in this paper 
generalize those constructed in Ref.~\cite{klimov1} for an $su_3$ three-level system (corresponding 
to $d=3$)~; they give rise to new phase states, namely, vector phase states which are eigenstates 
obtained along lines similar to those developed in Ref.~\cite{twareque1, twareque3} 
for obtaining a vectorial generalization of the coherent states introduced in Ref.~\cite{gazeau}. Still 
for $\kappa < 0$, a new type of unitary phase operator ($E_d$) can be defined~; it specificity is to 
span all vectors of the $d$-dimensional representation space of ${\cal A}_{\kappa}(2)$ from any vector of the 
space.  

In the case $\kappa \geq 0$, it is possible to define nonunitary phase operators. They can be turned to 
unitary phase operators by truncating (to some finite but arbitrarily large order) the representation 
space of ${\cal A}_{\kappa}(2)$. This leads to a truncated generalized oscillator algebra (${\cal A}_{\kappa, \sigma}(2)$) 
that can be reduced to the Pegg-Barnett truncated oscillator algebra$^{\cite{pegg-barnett}}$ through an 
appropriate limiting process where $\kappa \to 0$. 

Among the various properties of the phase states and vector phase states derived for $\kappa < 0$ and 
$\kappa \geq 0$, the property of temporal stability is essential. It has no equivalent in Ref.~\cite{Vourdas1990}. In last analysis, this 
property results from the introduction of a phase factor ($\varphi$) in the action of the annihilation and creation 
operators of ${\cal A}_{\kappa}(2)$. As an unexpected result, the quantization of this phase factor allows to derive 
mutually unbiased bases from temporally stable phase states for $\kappa  < 0$. This is a further evidence that 
``phases do matters after all''$^{\cite{Klimov06}}$ and are important in quantum mechanics. 

\section*{ACKNOWLEDGMENTS}

MD would like to thank the hospitality and kindness of the 
{\em Service de physique th\'eorique de l'Institut de 
Physique Nucl\'eaire de Lyon} where this work was done.

\end{document}